\documentclass[twocolumn,prl,superscriptaddress,showpacs,preprintnumbers,amsmath,amssymb]{revtex4-1}
\usepackage{graphicx}
\usepackage{color}
\usepackage{cases}
\usepackage{ulem}

\renewcommand{\vec}{\bf }

\begin{document}

\title{How does a thermal binary crystal break under shear?}
\author{Tobias Horn}
\email{thorn@thphy.uni-duesseldorf.de}
\author{Hartmut L\"owen}
\affiliation{Institut f\"ur Theoretische Physik II: Weiche Materie, Heinrich-Heine-Universit\"at D\"usseldorf,
D-40225 D\"usseldorf, Germany}
\date{\today}

\begin{abstract}
When exposed to strong shearing, the particles in a crystal will rearrange and ultimately, the crystal will break by forming
large nonaffine defects. Even for the initial stage of this process, only little effort has been devoted
to the understanding of the breaking process on the scale of the individual particle size
for thermalized mixed crystals. Here, we explore the shear-induced breaking for an equimolar
 two-dimensional binary model crystal with a high interaction asymmetry between the two different species such that the initial crystal has 
an intersecting square sublattice of the two constituents. Using Brownian dynamics computer simulations, we show
that the combination of shear and thermal fluctuations leads to a characteristic hierarchical breaking scenario where initially, the
more strongly coupled particles are thermally distorted, paving the way for the weakly coupled particles to escape from their cage.
This in turn leads to mobile defects which may finally merge, proliferating a cascade of defects,
which triggers the final breakage of the crystal. This scenario is in marked contrast to the breakage of one-component crystals close to melting.
Moreover, we explore the orientational dependence of the initial shear direction relative to the
crystal orientation and compare this to the usual melting scenario without shear. Our results are verifiable in real-space
experiments of superparamagnetic colloidal mixtures at a pending air-water interface in an external magnetic field where the shear
can be induced by an external laser field.
\end{abstract}


\maketitle
%
%
\section{I. Introduction}\noindent
The stability of crystalline solids with respect to mechanical stress is a crucial property for the understanding of effects like microcrack formation and -propagation \cite{brau2014}, 
which have numerous applications ranging from aeronautic engineering \cite{alexopoulos2013,sang1995} to the biomechanics of bone fracture \cite{xe1998,najafi2007}. Thereby, shear deformation constitutes an elementary source of strain. Colloids pose an ideal model system for studying distortions on the particle scale \cite{ivlev2012} 
and allow to access the microscopic processes governing structural changes via experiment
\cite{baumgartl2007,heinrich2011,ackerson1981,keim2007,dullens2011,gerbode2010,mcmullan2010,pasquino2010,villanova-vidal2009, zanghellini2005, medebach2003,kaldasch1998,palberg1994,schall2007,deutschlaender2013,henseler2010} and computer simulation
\cite{butler1995,wilms2013,wilms2012,jaeger2013,smallenburg2012,kreuter2013,franzrahe2008,vezirov2013,Kruppa2012,derks2009}.
For one-component systems, the plastic deformation of a strained solid is well-explored and a connection between mesoscopic deformation and atomistic rearrangements has been established \cite{miguel2001,dimiduk2006,bulatov1998,vandermeer2014,shiba2010,hamanaka2008}. Many real solids, however, are multicomponent materials. Examples include metallic alloys \cite{fallah2012, fallah2013, rougier2013, sun2013, haghighat2013, mishin2013, matsumoto2013,iwai1999,yu2007} and crystalline organic networks \cite{ross2009,muto1989}. 
Therefore, it is important to study the behaviour of mixture  crystals under strong shear.
Even for the initial stage of the shear-induced breaking process of crystals, only little effort has been devoted
so far to understanding the underlying mechanism on the particle-resolved scale 
for thermalized mixed crystals.\\
Here, we explore the shear-induced breaking of an equimolar
 two-dimensional binary model crystal with a high interaction asymmetry between the two different species such that the initial crystal has 
an intersecting square sublattice of the two constituents. Using Brownian dynamics computer simulations, we show
that the combination of shear and thermal fluctuations leads to a characteristic hierarchical breaking scenario.
 In the strained unit cell, aligned displacements of the more strongly coupled particle species open up pathways for the motion of the weakly interacting particle species,
thus enabling the creation of vacancy/interstitial pairs.
These pathways correspond to a characteristic distortion of the energy landscape encountered by the weakly coupled species and depend on 
the crystal orientation with respect to the direction of shear. The microscopic mechanism inducing defect formation under shear can be distinguished from the mechanism governing
defect formation when the crystal is heated instead of sheared. Our results imply that the location of spontaneously created defects triggers the mesoscopic deformation of the crystal, i.e. the formation
of cracks, which is in striking contrast to the behaviour of one-component crystals near melting \cite{vandermeer2014}.
Furthermore, the breakage of the thermal binary crystal resembles the plastic deformation of amorphous materials \cite{varnik2003_jcp,shi2005,maloney2006,zausch2008,zausch2009,sentjabrskaja2013,picard2005,baret2002,zausch2009}, where,
due to the absence of distinct topological defects, plastic deformation is mediated by localized patterns of nonaffine motion \cite{maeda1981,tanguy2006,maloney2006,falk1998,yang2014,tondl2014,chen2012,desmond2013} and can be traced by the inherent stress
signature and spatial correlation of plastic events \cite{varnik2014,mandal2014_discussion,desmond2014} or contact force distributions \cite{boberski2014}. Our results are verifiable in real-space
experiments of superparamagnetic colloidal mixtures at a pending air-water interface in an external magnetic field \cite{kollmann2002,hoffmann2006,assoud2009,assoud2009jpcm,mishra2014}, where the shear
can be induced by an external laser field.\\
The paper is organized as follows: in section II, we describe the binary model crystal. The simulation technique is depicted in section III. 
Section IV contains a description of our diagnostics and analysis of defects. In section V, we discuss our results for two different shear directions relative to the
crystal orientation. We also compare our findings to the melting of the crystal in the absence of shear and to the shear response of a one-component crystal. Additionally, we discuss similarities to the plastic deformation of amorphous media. We conclude in section VI.
\section{II. Model}
\begin{figure}[t]
\centering
\includegraphics[width=.95\linewidth]{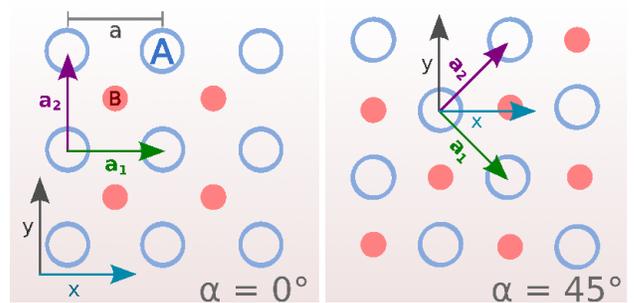}
\caption{Schematic representation of the S($AB$) lattice with a checkerboard structure of $A$ and $B$ particles. $A$($B$) particles are shown as blue open (red full) circles. The sublattices of both species
have a lattice constant $a$ while the two lattices are shifted relatively by $0.5a$ along the lattice vectors $\vec{a}_1$, $\vec{a}_2$. The direction of shear flow coincides with the $x$-axis.
The angle between $\vec{a}_1$ and the $x$-axis is denoted as $\alpha$. We consider the cases $\alpha = 0^{\circ}$ (left) and $\alpha = 45^{\circ}$ (right).  }
\label{fig_schematic}
\end{figure}

\noindent
We study the shear deformation of a two-dimensional colloidal crystal composed of two species of point-like particles denoted as species $A$ and $B$.
The particles interact via the purely repulsive pair potential of parallel dipoles and are characterized by different dipole moments $m_A$ and $m_B$, where
the dipolar ratio $m_B/m_A$ is fixed to 0.1 as in previous studies \cite{assoud2009,assoud2011}. Thus, particle species $A$ denotes the more strongly coupled particles. The crystal contains equal numbers of the two particle species, i.e., $N = N_A + N_B$ with a fixed relative composition $X = N_A/(N_A + N_B) = 0.5$.
In the absence of shear, the binary crystal lattice corresponds to the S($AB$) pattern specified in \cite{assoud2007}, where each particle species forms a quadratic lattice with spacing $a = 1/\sqrt{n_A}$, with $n_A$ denoting the number density of $A$ particles. The
lattices of species $A$ and $B$ are shifted relatively by $0.5a$ along each lattice direction, forming a checkerboard structure, see Fig. \ref{fig_schematic}. 
As elaborated in \cite{assoud2007}, the S($AB$) lattice is stable for a composition ratio of $X = 0.5$ and a 
dipolar ratio of 0.1.
This structure is very persistent in two dimensional mixtures and was found for granulates \cite{kaufman2009} and ionic crystals, as well \cite{assoud2010, assoud2008}.
In previous experiments, two-dimensional suspensions were studied by confining superparamagnetic colloidal particles to the air-water 
interface of a hanging water droplet \cite{zahn1999,Zahn2000,keim2007,assoud2009} or to a planar glass substrate \cite{deutschlaender2013,DeutschlaenderHorn2013,horn2013}. Parallely aligned dipole moments are induced by applying an external
magnetic field $H$ perpendicular to the plane of confinement.\\Conveniently, the pair interaction strength can be expressed by the dimensionless parameter 
\begin{equation}\label{gamma}
 \Gamma=\frac{\mu_0\left(\chi_A H\right)^2}{4 \pi a^3k_BT}\ ,
\end{equation}
where $\mu_0$ is the vacuum permeability, $\chi_A$ denotes the magnetic susceptibility of particle species $A$, $k_BT$ is the thermal energy and $a$ is the lattice constant as specified above. By altering the magnetic field, the colloidal suspension can be effectively ``heated'' or ``cooled'' homogeneously.\\
Accordingly, the species-dependent pair interaction potential is given in units of $k_BT$ by
\begin{equation}
 U_{ij}(r) = \Gamma\ m_i m_j (r)^{-3} \qquad i = A,B \quad,
\end{equation}
where $r$ denotes the particle separation in units of the lattice constant $a$. 
Unless otherwise stated, the interaction strength is set to $\Gamma = 44$ such that crystalline order is maintained in the absence of shear flow \cite{stability_remark}.
\section{III. Brownian Dynamics Simulation}\noindent
We perform Brownian dynamics (BD) computer simulations of $N = 2048$ pointlike superparamagnetic particles. The ratio of diffusion constants $D_B/D_A$ is fixed to 1.7 corresponding to the physical diameter ratio of comparable experimental samples \cite{assoud2009}.
The Brownian time of $A$ particles, $\tau_B = a^2/D_A$ is employed as a unit of time, distances are given in units of the lattice constant $a$.
The particles are initially placed on an S($AB$) lattice as described in section II, which covers the entire simulation area $L_x \times L_y = 32a \times 32a$.
Couette shear flow is imposed in $x$-direction at a fixed shear rate $\dot{\gamma}\tau_B = 10^{-3}$ and Lees-Edwards boundary conditions are applied according to the shear flow \cite{lees1972}. 
The equation of motion governing the particle propagation is given by
\begin{align}
	   x(t+\delta t) & = x(t) + \frac{D_i}{k_BT}F^{(x)} \delta t + \sqrt{2D_i\delta t}{R^{(x)}}\nonumber\\
	   & + \dot{\gamma}(y)\delta t, \nonumber \\
	   y(t+\delta t) & = y(t) + \frac{D_i}{k_BT}F^{(y)} \delta t + \sqrt{2D_i\delta t}{R^{(y)}},\nonumber\\
	   \qquad i = A,B\nonumber
\end{align}
where $F$ denotes the force resulting from the particle interaction and $R^{(x)},R^{(y)}$ are randomly distributed numbers with mean 0 and standard deviation 1.
The incremental timestep is set to $\delta t / \tau_B = 10^{-5}$. 
Forces are truncated and shifted according to
\begin{equation}
	F_{ij}(r) =
	\begin{cases}
	-3\Gamma m_im_j \left(\frac{1}{r^{4}} - \frac{1}{r_c^{4}}\right) & r < r_c\\
	0 & r \geq r_c\\
	\end{cases}
\end{equation}
where the cutoff is set to $r_c/a = 8$. Simulations are performed over a time range of $t = 200 \tau_B$ equaling a total strain window of $\gamma = \dot{\gamma}t = 0.2$, 
while configurations are recorded every $10^{-2}\tau_B$. The shear deformation process of the binary S($AB$) crystal is analyzed for two different crystal orientations with respect to the direction of shear.
In the first setup, the crystal is aligned in such a way that the crystal axis $\vec{a}_1$ collapses with the direction of shear, $\alpha = \angle(\vec{a}_1,\vec{e}_x) = 0^{\circ}$. 
In the second setup, the crystal is rotated such that $\alpha = 45^{\circ}$, see Figure \ref{fig_schematic}. Note that in this setup, we chose $N = 2304$ and $L_x = L_y = 24\cdot\sqrt{2}a$ in order to accomodate the rotated crystal. 
Each case is sampled with 64 statistically indepent simulation runs. 
In each run, the initial pure, defect-free crystalline lattice is allowed to relax for $10\tau_B$ before shear is applied. 
Note that due to the focus on a defect-free initial state, we discard runs in which topological defects emerge during the initial relaxation phase. This was the case in approximately $8\%$ of the simulation runs.
\section{IV. Diagnostics and analysis of defects}
\begin{figure}[t]
\centering
\includegraphics[width=.95\linewidth]{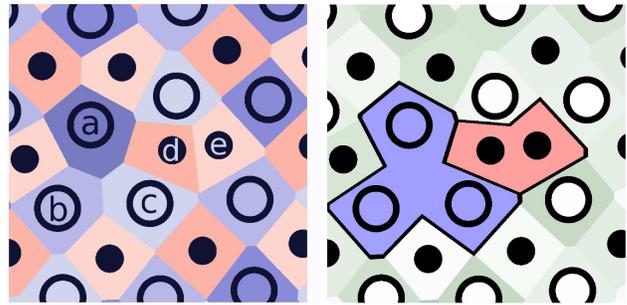}
\caption{Computer simulation snapshots exemplifying the definition of clusters based on Voronoi tesselation. $A$($B$) particles are represented by open (full) circles.
Left: Neighboring $A$ particles \textbf{a},\textbf{b} and \textbf{c} each share one neighbor of particle species $B$ and thus form a cluster. Also, $B$ particles \textbf{d} and \textbf{e} form a cluster. Note that particles \textbf{e} has less than four neighbors of species $A$ while particle \textbf{d} has four $A$ neighbors here 
but is considered part of the cluster due to the subthreshold distance to cluster particle \textbf{e}. 
Voronoi cells of $A$($B$) particles are colored in shades of blue (red). Right: Resulting cluster configuration, where Voronoi cells of $A$($B$) particles forming a cluster are colored in blue (red).}
\label{fig_instructive}
\end{figure}

\begin{figure}[t]
\centering
\includegraphics[width=.95\linewidth]{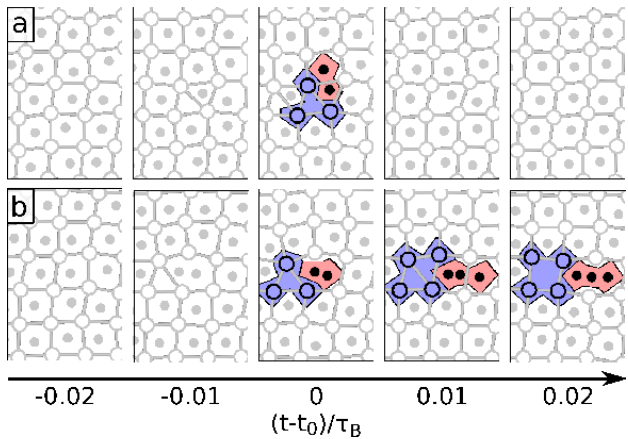}
\caption{Computer simulation snapshots illustrating the formation of a temporary (a) and a persistent (b) cluster pair at the time $t_0$ due to thermal fluctuations. From left to right, each row corresponds to a series
of snapshots recorded in intervals of $10^{-2}\tau_B$. 
The two events shown were recorded in different simulation runs and are uncorrelated. The representation of particles and clusters corresponds to Fig. \ref{fig_instructive}. As a guide to the eye, bonds between adjacent $A$ particles are shown if the seperation is less than $1.2a$.}
\label{fig_early}
\end{figure}

\noindent
In order to track structural distortions over various stages of the deformation process, we devise a cluster criterion, by which topological distortions are detected
based on the Voronoi tesselation of a particular configuration.
We consider only clusters which are constituted exclusively by $A$ or $B$ particles, respectively.
In a perfect S($AB$) lattice, every $A$ particle has four $A$ neighbors, each of which shares two $B$ neighbors with the original particle. 
Neighboring $A$ particles form a cluster if they share less than two neighbors of species $B$.
In the ideal S($AB$) lattice, every $B$ particle is enclosed by four neigbors of the opposite species. Neighboring $B$ particles form a cluster if each of them has less than four $A$ neighbors.
To grant the proper representation of interstitials, where two $B$ particles occupy the same lattice site, a $B$ particle is also counted as part of the cluster if it has four $A$ neighbors but is within a critical distance to a $B$ particle with less than four
$A$ neighbors \cite{cluster_remark}.  For both particle species, clusters are discarded if they contain less than two particles. 
For clarity, the definition of $A$ and $B$ clusters stated above is exemplified in Fig. \ref{fig_instructive}.
The clusters defined by these criteria pose elementary topological distortions of the S($AB$) lattice. Specifically, the local appearance of an $A$ and a $B$ cluster constitutes a primary plastic event which compromises the crystalline structure and may irreversibly
distort the crystal by evolving into a stable vacancy/interstitial pair, see Fig. \ref{fig_early}.\\
In order to quantify the lifespan of defects and local distortions, a cluster $\mathcal{C}$ detected in configuration $\mathcal{N} = (\vec{r}^A_1,..,\vec{r}^A_{N_A},\vec{r}^B_1,..,\vec{r}^B_{N_B})$ is considered 
identical to the cluster $\mathcal{C^{\prime}}$ detected in the preceding configuration $\mathcal{N}-1$ if at least one of the particles constituting $\mathcal{C^{\prime}}$ is conserved in $\mathcal{C}$, no other cluster in the current configuration $\mathcal{N}$
contains more particles from $\mathcal{C^{\prime}}$ than $\mathcal{C}$ and more particles in $\mathcal{C}$ originate from $\mathcal{C^{\prime}}$ than from any other previous cluster. 
Thus, we account for fluctuations in the cluster size and composition, cluster recombinations or splits while keeping track of both localized and mobile structural deformations. Since the time interval
between two successively recorded configurations is $10^{-2}\tau_B$, this value is set as the lifespan of a cluster which spontaneously emerges and vanishes again in the next configuration recorded.\\
\noindent Based on this cluster description, a pattern is established to capture spontaneous attempts to create vacancy/intersitial pairs. These events are detected if they bear the following signature:
In configuration $\mathcal{N}$, a cluster $\mathcal{C}_A$ is detected comprising three or more particles of species $A$. Additionally, a cluster $\mathcal{C}_B$ is detected which is composed of at least two $B$ particles.
In the last recorded configuration $\mathcal{N}-1$, $\mathcal{C}_A$ and $\mathcal{C}_B$ did not exist and none of the particles constituting $\mathcal{C}_A$ and $\mathcal{C}_B$ were part of any cluster. Additionally, in configuration $\mathcal{N}-1$, at least one of the $B$ particles in
$\mathcal{C}_B$ was a Voronoi neighbor of at least 3 of the $A$ particles forming $\mathcal{C}_A$. This particle is denoted the trigger particle.\\
The stress induced by the shear flow in $x$-direction is expressed via the stress tensor component $\sigma_{xy}$ \cite{heyes1994}:
\begin{equation}\label{sigma_xy}
	\sigma_{xy}=\frac{1}{2}\frac{1}{L_xL_y}\sum_{i}\sum_{i \neq j}F^y_{ij}(r_{ij})x_{ij}\ ,
\end{equation}
where $F^y_{ij}$ denotes the $y$-component of the dipolar force between particles $i$ and $j$ and $r_{ij}$ ($x_{ij}$) is the (lateral) distance between the particles in units of the lattice constant $a$.
\section{V. Results}\noindent
\subsection{1. Shear aligned with lattice direction}\noindent
In the following, we depict the shear deformation process observed for the case $\alpha = 0^{\circ}$, i.e. for the shear aligned with the lattice direction.\\
In the \textit{initial stage} of the deformation process, the crystal responds elastically to the applied shear and particle motion is governed by a uniform affine motion corresponding to the imposed solvent flow. Locally, 
we observe the formation of short-lived clusters with lifespans of the order $10^{-2}\tau_B$. 
Particle-scale observations suggest that these events are induced by thermal fluctuations of the $A$ particle species: 
Deviating from their co-sheared lattice site, single $A$ particles temporarily distort the unit cell symmetry, thus allowing the enclosed $B$ particle to extend its motion
into the vacated area and to form a cluster with the $B$ particle of the neighboring unit cell. Simultaneously, the remaining three $A$ particles constituting the original unit cell shift into an unstable triangular arrangement, and a cluster is formed.
Starting from a local configuration without any noticeable distortion, these structural disruptions emerge and disappear within a few $10^{-2}\tau_B$.
After this time, the original shape of the unit cell is restored. An exemplary sequence of simulation snapshots illustrating a temporary cluster creation due to thermal fluctuations is shown in Fig. \ref{fig_early}a.
\begin{figure}[t]
  \includegraphics[width=.95\linewidth]{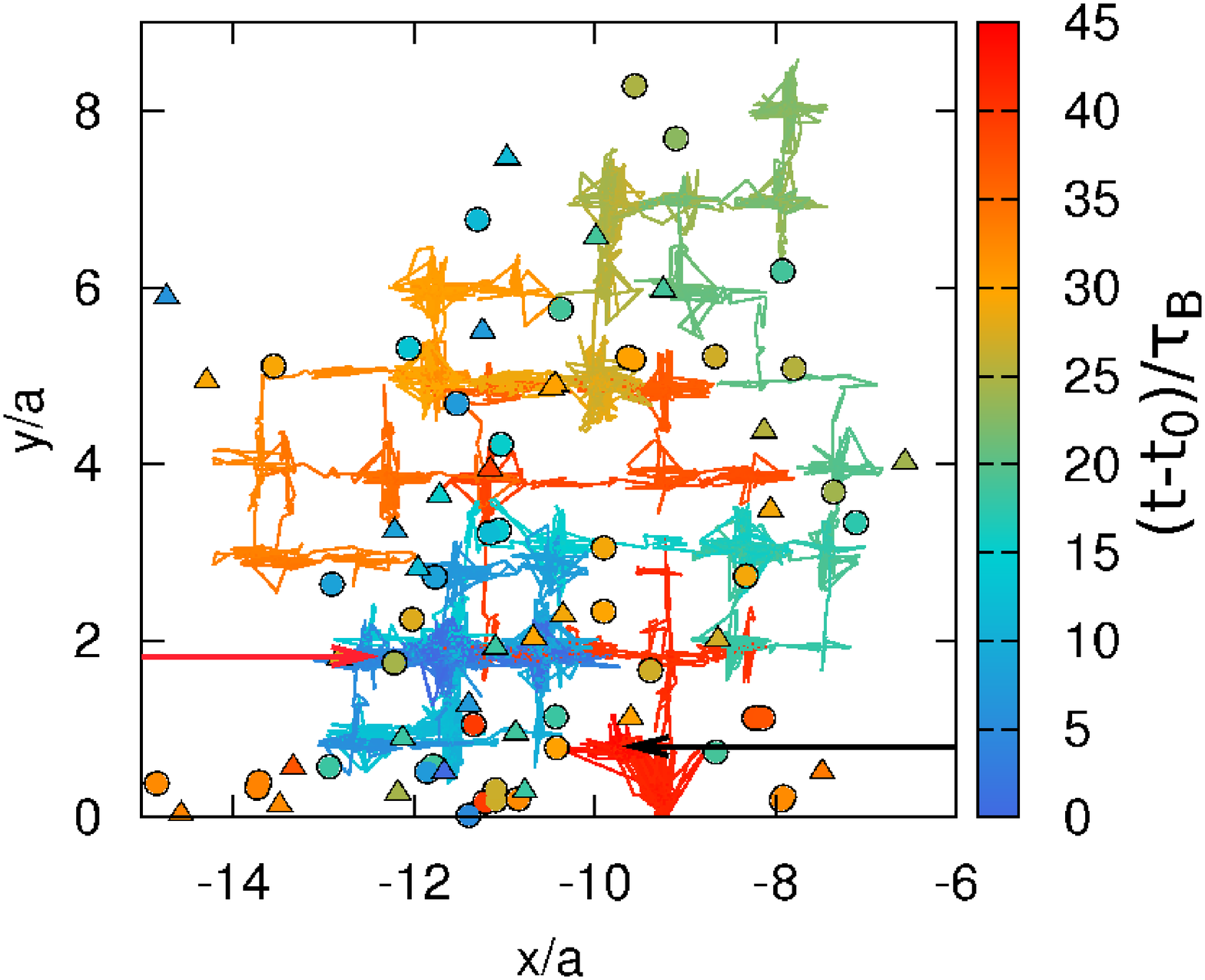}
  \caption{Trajectory of a single $B$ particle cluster from creation at $t=t_0$ (red arrow) to annihilation at $t=t_1$ (black arrow). The cluster emerges in the second stage of the deformation process, i.e. after a stable vacancy/interstitial pair was formed.
  The colorcode corresponds to the lifetime of the cluster. Circles (triangles) indicate the location of subsequent $A$($B$) cluster appearances within the lifetime of the travelling cluster.}
\label{fig_interstitial}
\end{figure}
\begin{figure*}[ht]
  \includegraphics[width=\textwidth]{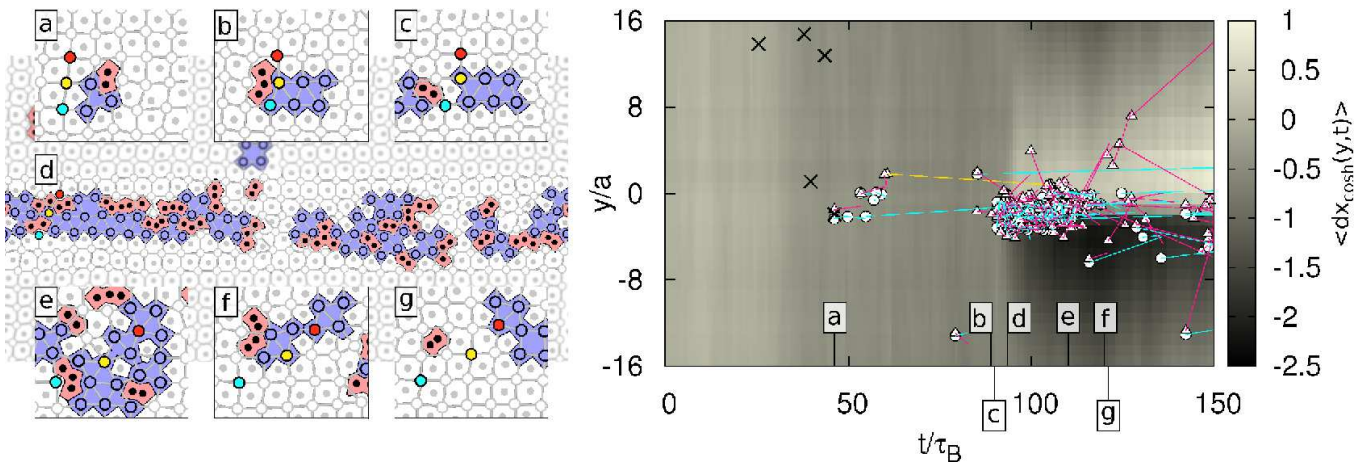}
  \caption{Left: Simulation snapshots illustrating the deformation process for the case $\alpha = 0^{\circ}$. All snapshots originate from the same simulation run.
Representation of particles and clusters corresponds to Fig. \ref{fig_instructive}. As a guide to the eye, three individual $A$ particles are shown in red, yellow and cyan, respectively.
a: Creation of stable cluster pair, b: extension of an $A$ cluster in $x$-direction, c: local rearrangement induced by interplay of $A$ and $B$ clusters, d: formation of an extended, disordered cluster strip and e: crystal
rearrangement on either side of the strip. f-g: Restoration of S($AB$) structure after nonaffine displacement on either side of strip (note shift of tagged particles). Right: Map of cumulative displacement in $x$-direction in the co-sheared frame. The data were averaged over the initial $x$-coordinate. Colors are
specified by the bar on the right. Black crosses indicate time and $y$-coordinate of short-lived clustering events. The appearance of $A$($B$) clusters with a lifetime $>0.5\tau_B$ is indicated with white full circles (triangles), straight cyan (pink) lines connect sites of creation and annihilation.
A yellow line indicates the lifespan of the cluster traced in Fig. \ref{fig_interstitial}. Times corresponding to snapshots a-g in left panel are indicated on the time axis.}
\label{fig_break}
\end{figure*}
\begin{figure}[hb]
\centering
\includegraphics[width=.95\linewidth]{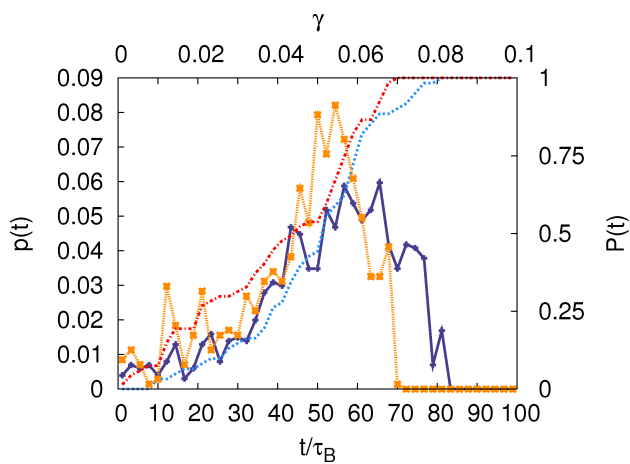}
\caption{Relative probability $p(t)$ for an event to occur at time $t$ for $\alpha = 0^{\circ}$ (blue solid line) and $\alpha = 45^{\circ}$ (yellow dashed line). The cumulative probability $P(t)$ for the 
appearance of stable defect up to the time $t$  is included for $\alpha = 0^{\circ}$ (light blue dashed line) and $\alpha = 45^{\circ}$ (red dashed line). For comparison, the absolute strain is given at the upper horizontal axis. The evaluation
is based on a total number of 1002 (706) recorded events for $\alpha = 0^{\circ}$ ($45^{\circ}$).}
\label{fig_event_time}
\end{figure}
\newline
The \textit{second stage} of the deformation process is initiated by the creation of a stable pair of clusters with a lifespan of the order $\tau_B$. 
Qualitatively, the particle motion leading to a persistent defect corresponds to the same pattern depicted above.
In addition to the formation of a triangular cluster of $A$ particles, the enclosed $B$ particle is pushed entirely out of the unit cell, enabling the formation of a vacancy/interstitial pair, i.e. a cluster
of four $A$ particles which assume a quadratic arrangement around a vacated $B$ lattice site and a cluster including two $B$ particles which occupy the same lattice site  (Fig. \ref{fig_early}b). The $B$ particle cluster created in this process
quickly diffuses through the crystal lattice, as the particles constituting the cluster are frequently interchanged. 
This process is mediated by $B$ particles pushing and replacing each other at the lattice sites touched by the travelling cluster.
The diffusive motion of the interstitial cluster does not exhibit a preferential direction and involves frequent reversals of the moving direction, see Fig. \ref{fig_interstitial}. Additionally, the travelling interstitial triggers the
spontaneous creation of further clusters at the lattice sites it touches. In contrast, location and particle constellation of the $A$ particle cluster formed in this event are comparably stable.\\
The \textit{third stage} of the deformation process starts with the formation of a larger cluster of $A$ particles. Being unstable with
respect to shear deformation, an elongated cluster of $A$ particles poses a preferential site for a break as it allows for a local rearrangement of the crystal lattice on either side of the cluster.
Once such an elongated cluster emerges, it quickly grows into a band spanning the entire system, which enables inhomogeneous, nonaffine particle displacements on the order of the lattice constant.
This pattern of motion resembles a shear band \cite{varnik2003,shi2005,park2007,schall2010,moorcroft2011,chaudhuri2012} but may also exhibit vortical properties as reported in \cite{maeda1981,tanguy2006,maloney2006}. 
Thus, the crystal is globally rearranged and a less strained state is restored. Intriguingly, particles in the proximity of the break are rearranged during the cluster formation
until eventually, the original S($AB$) structure can be recovered for most particles, although typically, some topological defects remain.\\
The deformation process depicted here is illustrated by computer simulation snapshots and a map of the cumulative nonaffine displacement in $x$-direction in Fig. \ref{fig_break}. 
The spontaneous emergence of
short-lived clusters can be observed at several positions as indicated by black crosses in Fig. \ref{fig_break} (right). Stable clusters emerge at $t/\tau_B \approx 50$ (white symbols), leading
to structural defects which persist over many $\tau_B$. The lifespan of stable clusters can be traced by straight lines in Fig. \ref{fig_break} (right). At  $t/\tau_B \approx 100$, we observe a strongly heterogeneous behavior of the cumulative displacement in $x$-direction
with respect to the $y$-position of particles: Above a certain $y$-position, particles undergo a pronounced nonaffine displacement in the direction of shear (bright area in Fig. \ref{fig_break} [right]). The nonaffine
cumulative $x$-displacement of particles below this $y$-position is oppositely directed (dark area in Fig. \ref{fig_break} [right]). This pattern of nonaffine motion reflects the breakage of the crystal and is accompanied by the creation and
annihilation of large numbers of clusters as the crystal rearranges. The location of the break coincides with the location of the first set of stable defects.
Note that the orientation of the break does not necessarily collapse with the direction of shear. We observe a vertical breakage of the crystal with a similar probability.\\\\
\begin{figure}[t]
\centering
\includegraphics[width=.95\linewidth]{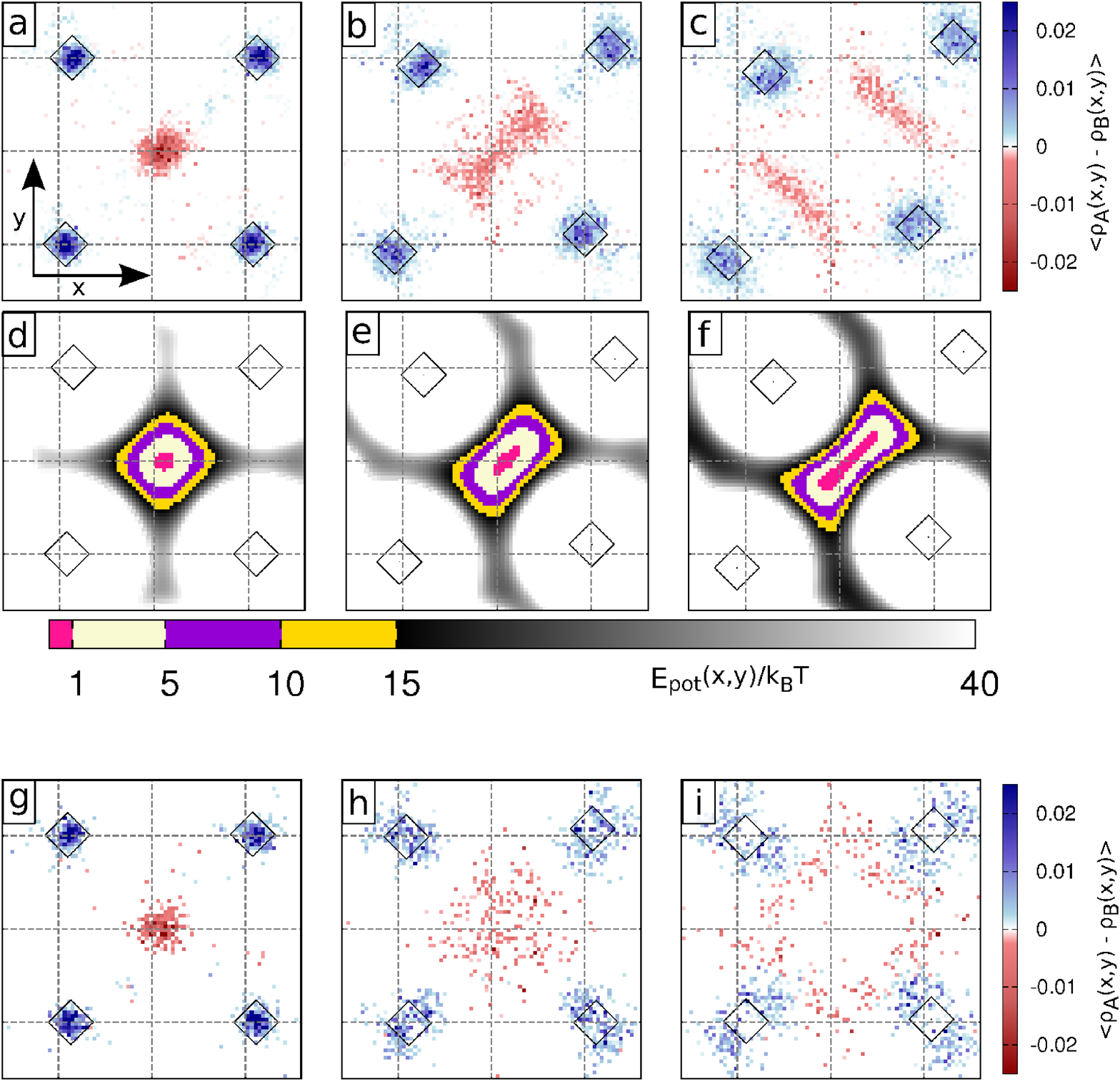}
\caption{$\alpha = 0^{\circ}$. a-c: Averaged densities $\langle\rho_A\rangle$, $\langle\rho_B\rangle$ of $A$ and $B$ particles at the site of an event at $t_0$ for times (a) $t-t_0 = -0.1\tau_B$, (b) $t-t_0 = -0.01\tau_B$ (c) $t-t_0 = 0$.
Colors correspond to $\langle\rho_A(x,y)\rangle - \langle\rho_B(x,y)\rangle$ and are specified by the bar on the right. Diamonds indicate the average position of $A$ particles.
d-f: Potential energy of a $B$ particle in the vicinity of $A$-particles located at the average positions in situations a-c. Colors are specified by the bar at the bottom. Potential energy is given
with respect to the mean position of the $B$ particle. g-i: Averaged densities preceeding a cluster creation when the crystal is heated to $\Gamma = 37$ and no shear flow is imposed ($\dot{\gamma}\tau_B = 0$).}
\label{fig_scenario_00}
\end{figure}
Since the emergence of a stable vacancy/interstitial pair in the course of a thermal clustering event crucially influences the macroscopic breakage of the crystal,
we thoroughly assess the properties of these events. In order to do so, we evaluate all events up to and including the first event which induces a cluster with a lifetime exceeding $0.5\tau_B$, which
corresponds to a total number of 1002 recorded events. Thus, we focus on the onset of deformation in a strained crystalline state with no embedded topological defects, i.e., the \textit{first stage} of the deformation process.  
We observe that within statistical accuracy, events are homogeneously distributed
over the entire system area and no preferential location can be distinguished.\\
As the absolute strain $\gamma$ grows, the probability of clustering events increases, see Fig. \ref{fig_event_time}. 
We derive a measure for the crystal stability with respect to shear by monitoring the integrated probability for the
creation of a long-lived cluster up to a given time. This magnitude is almost unity for times larger than $t/\tau_B \approx 80$, corresponding to an absolute strain of $\gamma \approx 0.08$.
This reflects the fact that in all of our simulation runs, a stable cluster has emerged up to this time, initiating the \textit{second stage} of the crystal deformation.\\\\
On a microscopic level, the collective particle motion leading to clustering exhibits a vast variety of patterns. In order to assess the common microscopic mechanism governing these events,
we monitor the average distribution of particles at the site of an event at three different times: $10^{-1}\tau_B$ before the event, $10^{-2}\tau_B$ before the event and at the
time of the event. The resulting density distributions of $A$ and $B$ particles are shown in Fig. \ref{fig_scenario_00}.
It is apparent from these distributions that events are permitted by a specific distortion of the unit cell, which originates from a superposition of two independent deformations.
First, the unit cell is strained due to a shear deformation along the $x$-direction, which collapses with the lattice vector $\vec{a}_1$. Second, the unit cell is distorted by a further
shear deformation of the $A$ particles along the second lattice direction, $\vec{a}_2$. This corresponds to a collective sliding motion of the $A$ particles on one side of the unit cell along the second crystal axis,
while the remaining $A$ particles collectively move into the opposite direction.\\
\begin{figure}[t]
  \includegraphics[width=.95\linewidth]{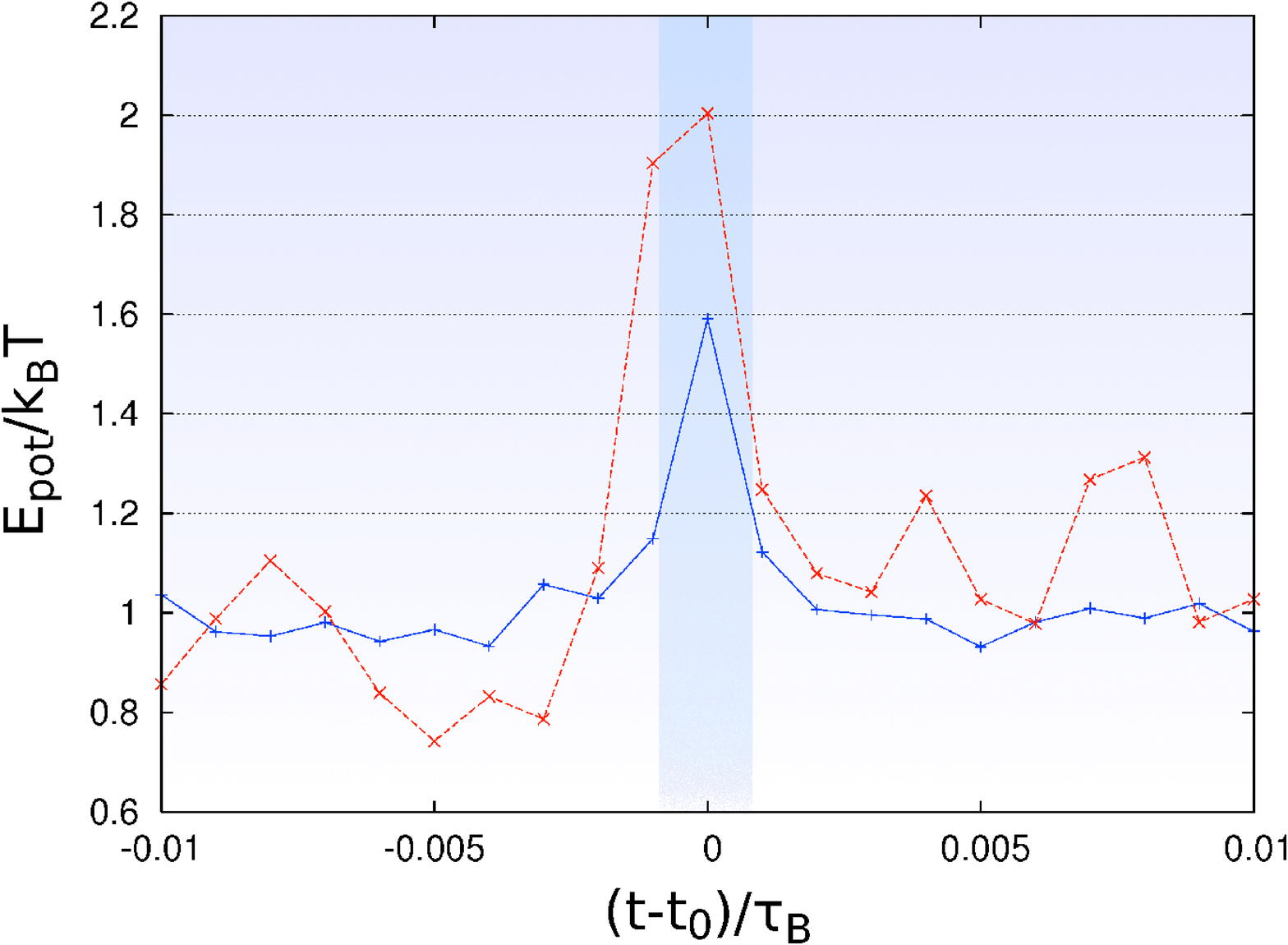}
  \caption{Potential energy of the trigger particle with respect to the closest energy minimum for an event at $t=t_0$ (highlighted region) for $\alpha = 0^{\circ}$. The blue solid curve corresponds to the average over all temporary events while the red dashed curve represents the average over all persistent events.}
\label{fig_thermal}
\end{figure}
Following this observation, we derive the average $A$ particle positions from the density distributions depicted in Fig. \ref{fig_scenario_00} (top) and map the potential energy landscape 
encountered by a single $B$ particle in the vicinity of four $A$ particles at the specified positions, see Fig. \ref{fig_scenario_00} (middle). In the reference state at $10^{-1}\tau_B$ before the event,
the average unit cell is only slightly sheared. There is a nearly quadratic energy well in the center of the unit cell to which the $B$ particle is confined. Due to the subsequent deformation of
the unit cell, the size of this potential energy well is reduced as it assumes a thin, tilted shape. Consequently, the $B$ particle is confined to a diagonally elongated region at $10^{-2}\tau_B$ before the event. 
In the following timestep, only a narrow potential energy well remains inside the unit cell due to the further distortion of the unit cell.
After the previous diagonal deflection of the $B$ particle, it is now pushed away from the center as the local energy minimum is displaced along four possible pathways.\\
From this shifted position, the
$B$ particle may escape the unit cell due to thermal motion. To verify the thermal activation of this event, we calculate the potential energy of the $B$ particle with respect to the nearest potential energy minimum. Indeed, the occurence of an event
is accompanied by a distinct peak of the potential energy on the order of $\approx 2k_BT$. Furthermore, we observe that the creation of a persistent cluster is related
to a significantly larger peak in the potential energy of the trigger particle than the creation of a temporary cluster, see Fig. \ref{fig_thermal}.
\newline
\newline
Thus, the investigation of particle density profiles at the location of events implies an intuitive picture for the formation of clusters from the strained crystalline state, 
which is permitted by two independent deformations of the unit cell: The global shear deformation and an aligned sliding motion of $A$ particles along the lattice direction $\vec{a}_2$, which does not
collapse with the direction of shear. In order to verify this explanation, we perform a set of reference simulations in which a single $B$ particle is placed in a distorted unit cell of four $A$ particles.
\begin{figure}[t]
\centering
\includegraphics[width=.95\linewidth]{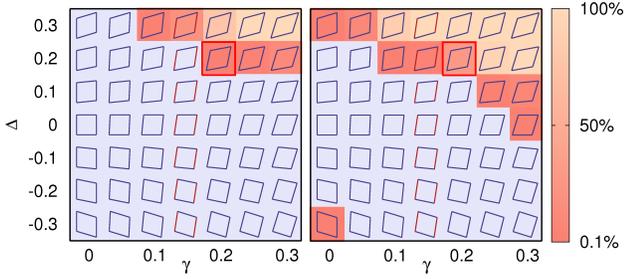}
\caption{Reference simulation for $\alpha = 0^{\circ}$, where a $B$ particle is initially trapped in a distorted unit cell formed by four $A$ particles, see text. The distortion is specified by $\gamma$ (shear deformation) and $\Delta$ (deflection
along the lattice vector $\vec{a}_2$, highlighted in dark red in the central column). Resulting unit cells are shown and 
fields are colorcoded according to the escape probability of a $B$ particle within a runtime of $1\ \tau_B$ (left) or $1000\ \tau_B$ (right),
based on 1000 samples per parameter combination. Escape probabilities below $0.1\%$ are shown in light blue. The parameter combination matching the average particle positons in Fig. \ref{fig_scenario_00}c is framed in red.}
\label{fig_trap_00}
\end{figure}
\begin{figure}
\centering
\includegraphics[width=.95\linewidth]{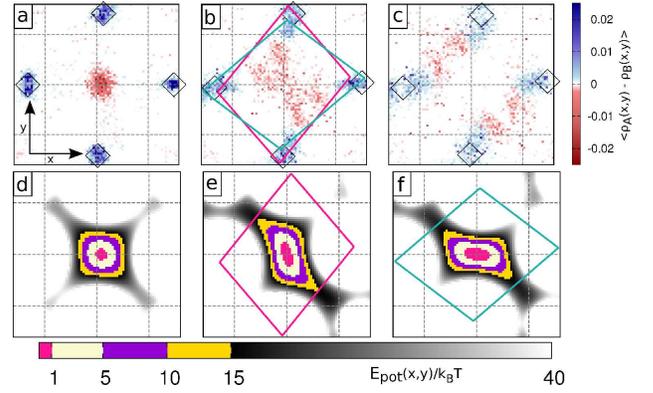}
\caption{$\alpha = 45^{\circ}$. a-c: Averaged local densities $\langle\rho_A\rangle$, $\langle\rho_B\rangle$ of $A$ and $B$ particles at the site of an event at $t_0$ for times (a) $t-t_0 = -0.1\tau_B$, (b) $t-t_0 = -0.01\tau_B$ (c) $t-t_0 = 0$.
Colors correspond to $\langle\rho_A(x,y)\rangle - \langle\rho_B(x,y)\rangle$ and are specified by the bar on the right. Diamonds indicate the average position of $A$ particles. For clarity, two distinct deformations
of the unit cell are included in panel b. d: Potential energy of a $B$ particle in the vicinity of $A$-particles located at the average positions in situation a. Panels e and f correspond to the unit cell outlines shown in b. Colors are specified by the bar at the bottom. Potential energy is given
with respect to the mean position of the $B$ particle.}
\label{fig_scenario_45}
\end{figure}
The distortion of the unit cell is expressed by the two independent parameters $\gamma$ and $\Delta$ : The parameter $\gamma$ corresponds to the unit cell strain in the direction of shear, while $\Delta$ 
quantifies the aligned deflection of the $A$ particles along the lattice direction $\vec{a}_2$, see Fig. \ref{fig_trap_00}. For each parameter combination, the $B$ particle trajectory is tracked
for $1\tau_B$ while the $A$ particles remain pinned to their distorted lattice positions. Each parameter combination is sampled with 1000 runs. Afterwards, we determine the ratio of runs in which
the $B$ particle escaped the unit cell. Our results are shown in Fig. \ref{fig_trap_00}. We find that, starting from a strained state, an additional deflection of the $A$ particles indeed leads
to an increased probability for the $B$ particle to escape the unit cell. The parameter combination of $\gamma$ and $\Delta$ which best matches the average particle positions derived from the $A$ particle density distribution is
highlighted in Fig. \ref{fig_trap_00}. In fact, this point in the $\gamma-\Delta$ plane coincides with the crossover from zero to a finite
value of the escape probability of the $B$ particle. This crossover is persistent if we extend the runtime of the simulation to $1000\tau_B$, as shown in Fig. \ref{fig_trap_00}.\\
\subsection{2. Shear unaligned with lattice direction}\noindent
In order to assess the dependence of the shear response on the direction of shear flow with respect to the crystal orientation, we repeat the analysis for the case $\alpha = 45^{\circ}$.\\
In comparison to $\alpha = 0^{\circ}$, we notice a more pronounced increase of the event probability with increasing strain.
Likewise, the integrated probability for the creation of a stable cluster up to a given time reaches unity at a time $t/\tau_B \approx 70$ ($\gamma \approx 0.07$),
see Fig. \ref{fig_event_time}. This points to a reduced stability of the crystal with respect to shear deformation when the unit
cell is rotated with respect to the direction of shear.\\
\begin{figure}[t]
\centering
\includegraphics[width=.95\linewidth]{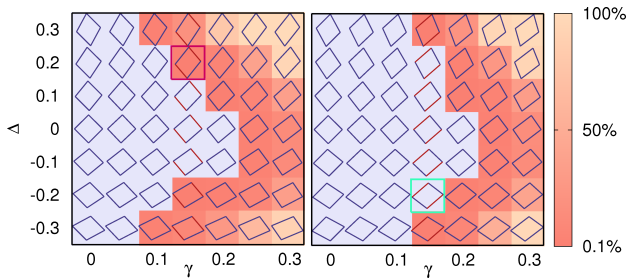}
\caption{Escape probability of a $B$ particle from a distorted unit cell for $\alpha = 45^{\circ}$. The distortion is specified by $\gamma$ (shear deformation) and $\Delta$ (deflection
along the lattice vectors $\vec{a}_1$ [left] or $\vec{a}_2$ [right], highlighted in dark red in the central column). Resulting unit cells are shown and fields are colorcoded according to the escape probability of a $B$ particle within a runtime of $1\ \tau_B$,
based on 1000 samples per parameter combination. Escape probabilities below $0.1\%$ are shown in light blue. The parameter combinations matching the cells shown in Fig. \ref{fig_scenario_00}b are framed.}
\label{fig_trap_45}
\end{figure}
Furthermore, we analyze the particle interplay leading to an event for $\alpha = 45^{\circ}$. Our results are shown in Fig. \ref{fig_scenario_45}.
As opposed to the previous case, the averaged density distributions do not point to a single deformation mode but rather depict the superposition of two distinct deformation modes, which
correspond to sliding motions of the $A$ particles along either lattice vector, $\vec{a}_1$ or $\vec{a}_2$. Since neither of the two lattice directions collapses with the
direction of shear, sliding motions along both directions distort the unit cell in an equivalent way. To support this interpretation, shifted unit cells are included in Fig. \ref{fig_scenario_45}, 
which are deformed according to either of the two modes. The potential energy landscape exhibits two pathways for the $B$ particle for each deformation mode. 
The superposition of these pathways can be recovered in the average density distribution of $B$ particles, which exhibits a cross-shaped maximum at $10^{-2}\tau_B$ before the event.\\
As with the case $\alpha = 0^{\circ}$, we perform additional simulations to sample the escape probability of a single $B$ particle enclosed in a distorted unit cell, where
the distortion is expressed by the parameters $\gamma$ and $\Delta$, see \ref{fig_scenario_00}. The parameter combinations corresponding to the distorted unit cells shown in \ref{fig_scenario_00} are
highlighted in the diagram and are close to the crossover to a finite escape probability.
\subsection{3. Comparison with different deformation scenarios}
\begin{figure}[t]
  \includegraphics[width=.95\linewidth]{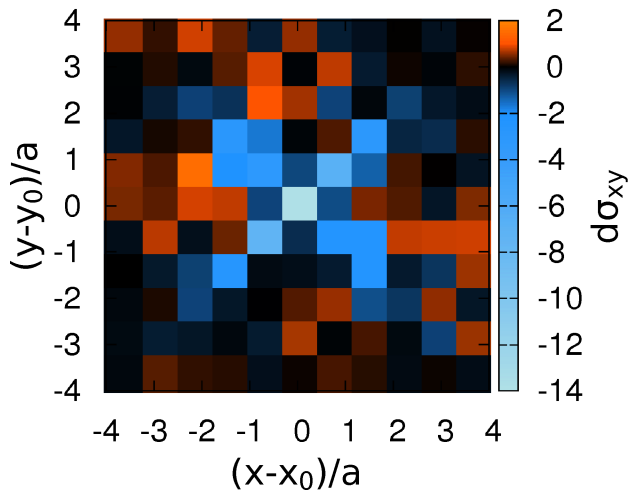}
  \caption{Averaged stress signature of events recorded for $\alpha = 0^{\circ}$ and $\alpha = 45^{\circ}$. The colorcode corresponds to the stress increase with respect to the previous configuration for particles
at a given distance to the trigger particle at $x_0$,$y_0$. The fourfold pattern qualitatively corresponds to recent findings for polydisperse hard spheres under shear \cite{mandal2014_discussion} and flowing emulsions near jamming \cite{desmond2014}.}
\label{fig_stress_signature}
\end{figure}

\begin{figure}[t]
  \includegraphics[width=.95\linewidth]{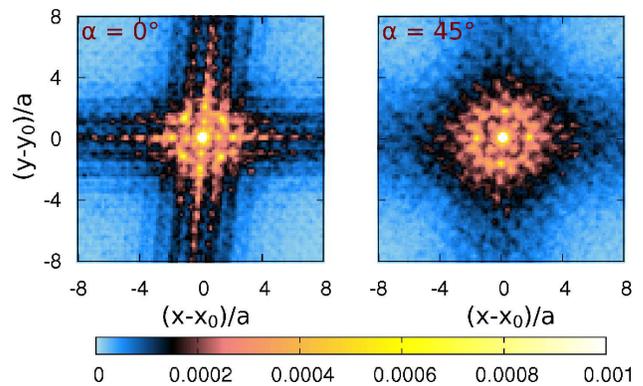}
 \caption{Spatial correlation of $A$ cluster appearances for $\alpha = 0^{\circ}$ (left) and  $\alpha = 45^{\circ}$ (left). The colorcode corresponds to the probability for a cluster to emerge at a particular relative position $x-x_0,y-y_0$ with
respect to a previous cluster creation at $x_0,y_0$. Colors are specified by the bar at the bottom.}
\label{fig_corr}
\end{figure}

\noindent
We compare the pattern of defect formation observed in the sheared binary crystal to the creation of defects when the crystal is heated to $\Gamma = 37$ and no shear flow is imposed ($\dot{\gamma} = 0$). 
Our results are included in Fig. \ref{fig_scenario_00} and indicate a qualitatively different mechanism with respect to the individual particle motion. Here, the crystalline structure cannot be attributed
to an aligned deflection of the $A$-particles but rather stems from a diffusion of the particles away from the unit cell center.\\\\
Furthermore, we qualitatively compare the deformation process of the sheared binary crystal to the shear deformation of a one-component crystal near melting. The crystal is constituted by $N = 2304$ dipolar particles
of species $A$ which are arranged in a triangular lattice spanning the simulation area $L_x \times L_y = 48a \times 41.57a$. The inverse temperature is set to $\Gamma = 20$ and
the shear rate is fixed at $\dot{\gamma} = 10^{-3}$ as before. For the one-component crystal, the structural rearrangment needed to release the strain is enabled by the spontaneous creation of (isolated) dislocation pairs along the tearing direction.
\begin{figure}[t]
\centering
\includegraphics[width=.95\linewidth]{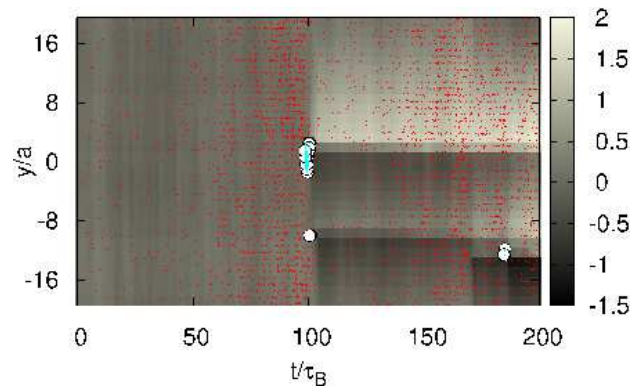}
\caption{Typical deformation process for the one-component crystal under shear. Colorcode corresponds to cumulative displacement in $x$-direction in the co-sheared frame. White circles indicate clusters with a lifetime above $0.1 \tau_B$, straight cyan lines connect sites of creation and annihilation. Short-lived
clusters with a lifetime below $0.1 \tau_B$ are indicated by red dots.}
\label{fig_break_mono}
\end{figure}
As opposed to the binary crystal, these topological defects can be spontaneously created to accomodate rearrangements of the adjacent particles. Thus, the onset of plasticity is not determined by the presence
of persistent topological defects, as is the case with the binary crystal. Similarly, the defects are quickly annihilated once the crystal rearranged into a less strained configuration. Thus, the defects observed in the sheared one-component
crystal are mostly short-lived with a lifetime well below $0.1 \tau_B$ and the deformation of the crystal does not entail an extended amorphous region. We observe that defects appear more frequently than in the
strained binary crystal. For a qualitative comparison, we devise a simplified cluster criterion for the one-component crystal, where
neighboring particles form a cluster if they each have either five or seven nearest neighbors, thus distorting the sixfold symmetry of the ideal lattice. 
Thereby, we cover isolated and bound dislocations as well as entangled chains. Figure \ref{fig_break_mono} depicts the creation and annihilation of clusters.\\
Recent studies on the plastic deformation of amorphous media have explicitly demonstrated a characteristic stress signature and spatial correlation of plastic events \cite{varnik2014,mandal2013,mandal2014_discussion,desmond2014},
corresponding to earlier theoretical predictions \cite{argon1979}. Encouraged by these findings, we assess the local stress signature of cluster creation events, which play the role of plastic events in the breaking of the S($AB$) crystal.
Intriguingly, our results are in qualitative agreement with the stress signature observed in an amorphous system of polydisperse hard spheres under shear \cite{mandal2014_discussion} and in flowing emulsions near jamming \cite{desmond2014}, see Fig. \ref{fig_stress_signature}. Additionally, we probe the spatial correlation of cluster
appearances and recover a non-vanishing spatial correlation along two distinct directions, which are aligned parallel and perpendicular to the direction of shear flow, respectively, see Fig. \ref{fig_corr}. This finding
is in agreement with the observation that the S($AB$) crystal may develop cracks in the lateral or vertical direction, as discussed above. A similar spatial correlation of defect creation sites was not observed for the reference
case of a sheared one-component crystal.
\section{VI. Conclusions}\noindent
We have analyzed the shear-induced breaking of a thermal binary model crystal on the particle-scale level using Brownian dynamics computer simulations. 
Thus, we have established that the macroscopic breakage of the crystal originates in the spontaneous creation of defects in the initially defect-free crystal. The defects are visualized by a cluster criterion.
Our observations suggest that the emergence of defects is induced by a characteristic hierarchical interplay of the
two particle species. In the strained crystal, aligned fluctuations of the more strongly interacting $A$ particles along the strained lattice directions distort the potential energy landscape and induce
plastic events, where the $B$ particles are pushed out of the unit cell. The likelihood of these events is increased if the shear direction is not aligned with the crystal orientation. 
Further plastic events are triggered by vacancy diffusion until eventually, small clusters merge into an extended disordered region and initiate the macroscopic breakage of the crystal,
during which pronounced nonaffine displacements are observed. This breaking scenario is in distinct contrast to the shear response of a one-component crystal near melting, where stress is released via the spontaneous
creation and annihilation of dislocation pairs without spatially extended or persistent regions of structural disorder. Furthermore, on a miscroscopic level, the breaking of the binary crystal under shear can be clearly distinguished from the melting of the crystal in the absence of shear.\\
Our observations suggest that various properties which are intrinsic to the plasticity of
amorphous materials can be found in the breaking of a binary crystal, where the escape of $B$ particles from their cages induces a local loss of structure and allows for nonaffine rearrangements in the manner of an amorphous material.
These properties include a fourfold stress signature of plastic events and a distinct, anisotropic spatial correlation of plastic events with respect to the shear direction.
Thus, our results shed new light on the study of solid plasticity, where crystalline and amorphous solids pose qualitatively different scenarios. Our results imply that in the case of a multicomponent crystal - even in the simple case of 
a binary crystal - this qualitative difference does not fully apply due to a local amorphization of the crystal in the breaking zone.\\
Our predictions can be verified in binary suspensions of superparamagnetic colloidal particles at a pending air-water interface in an external magnetic field
\cite{zahn1999,Zahn2000,gruenberg2004,keim2007}. The shear can be imposed by a laser beam.
For the future, one should explore more situations of shear-induced breaking. In particular,
the case of small damping will be interesting as this is relevant for the melting
of a dusty plasma crystal under shear (have a look at our book for references).
Moreover, more interaction asymmetries, composition ratios and model potentials need to be explored.
Last, the case of three spatial dimensions would be interesting where a wealth of
binary colloidal crystals are possible \cite{bartlett1992}. 
\section{Acknowledgments}\noindent
We thank J. F. Brady, S. Mandal, J. Horbach and P. Chaudhuri for valueable advice. We thank C. V. Achim for providing a computer code. This work
was supported by the SFF from the HHU.
\bibliographystyle{prsty}

\begin{thebibliography}{10}
\bibitem{stability_remark}
The value $\Gamma = 44$ was established as the liquid-solid coexistence value
  of the model system at hand in a previous simulation study by the authors.
\bibitem{cluster_remark}
The critical distance is set to 0.62$a$ but may be varied without qualitatively
  affecting our results.
\end{thebibliography}

\IfFileExists{\jobname.bbl}{}
 {\typeout{}
  \typeout{******************************************}
  \typeout{** Please run "bibtex \jobname" to optain}
  \typeout{** the bibliography and then re-run LaTeX}
  \typeout{** twice to fix the references!}
  \typeout{******************************************}
  \typeout{}
 }

\end{document}